\documentclass[aps,pra,twocolumn,showpacs,amsmath]{revtex4-1}
\usepackage{CJKutf8}
\usepackage{amsmath}
\usepackage{amsfonts}
\usepackage{amssymb}
\usepackage{siunitx}
\DeclareSIUnit\gauss{G}
\usepackage[english]{babel}
\usepackage{CJK}
\usepackage{rotating, graphicx}
\usepackage{subfigure}
\usepackage{dcolumn}
\usepackage{bm}
\usepackage{units}
\usepackage{color}
\usepackage{longtable}
\usepackage{array}
\usepackage{multirow}
\usepackage{tablefootnote}
\usepackage{romannum}
\usepackage{comment}
\usepackage{braket}
\usepackage[letterpaper,total={7in,9.5in},top=0.75in,left=0.75in]{geometry}
\usepackage[colorlinks=true,linkcolor=blue,citecolor=blue,urlcolor=blue,filecolor=blue]{hyperref}
\usepackage{bm}
\usepackage{hyperref}
\usepackage{graphicx}
\usepackage[english]{babel}
\usepackage[T1]{fontenc}
\usepackage{xcolor}
\usepackage{amssymb}
\usepackage{notes2bib}
\usepackage{natbib}

\newcolumntype{L}[1]{>{\raggedright\let\newline\\\arraybackslash\hspace{0pt}}m{#1}}
\newcolumntype{C}[1]{>{\centering\let\newline\\\arraybackslash\hspace{0pt}}m{#1}}
\newcolumntype{R}[1]{>{\raggedleft\let\newline\\\arraybackslash\hspace{0pt}}m{#1}}

\graphicspath {{figures/}}

\begin{document}

\author{Oleksiy Onishchenko}
\altaffiliation[present address: ]{LaserLaB, Department of Physics and Astronomy, VU University, De Boelelaan 1081, 1081 HV Amsterdam, The Netherlands}
\email{1S03P2@strontiumBEC.com}
\affiliation{Van  der  Waals-Zeeman  Institute,  Institute  of  Physics,  University  of Amsterdam, Science  Park  904,  1098  XH  Amsterdam,  The  Netherlands}
\author{Sergey Pyatchenkov}
\author{Alexander Urech}
\author{Chun-Chia Chen (陳俊嘉)}
\author{Shayne Bennetts}
\author{Georgios A. Siviloglou}
\altaffiliation[present address: ]{Shenzhen Institute for Quantum Science and Engineering,  and Department of Physics, Southern University of Science and Technology, Shenzhen, 518055, People’s Republic of China}
\author{Florian Schreck}
\affiliation{Van  der  Waals-Zeeman  Institute,  Institute  of  Physics,  University  of Amsterdam, Science  Park  904,  1098  XH  Amsterdam,  The  Netherlands}

\title{The frequency of the ultranarrow ${^1\text{S}_0} - {^3\text{P}_2}$ transition in $^{87}\text{Sr}$}

\date{\today}

\begin{abstract}
We determine the frequency of the ultranarrow $^{87}\text{Sr}$ ${^{1}\text{S}_{0}} - {^{3}\text{P}_{2}}$ transition by spectroscopy of an ultracold gas. This transition is referenced to four molecular iodine lines that are observed by Doppler-free saturation spectroscopy of hot iodine vapor. The frequency differences between the Sr and the I$_2$ transitions are measured with an uncertainty of \SI{250}{\kilo\hertz}. The absolute frequency of the $^{87}\text{Sr}$ ${^{1}\text{S}_{0}} - {^{3}\text{P}_{2}}$ ($\text{F}'=7/2$) transition is $\SI{446648775(30)}{\mega\hertz}$ and limited in accuracy by the iodine reference. This work prepares the use of the Sr ${^{1}\text{S}_{0}} - {^{3}\text{P}_{2}}$ transition for quantum simulation and computation.
\end{abstract}

\begin{CJK*}{UTF8}{min}
\maketitle
\end{CJK*}

\pagenumbering{arabic}

\section{Introduction}

Atoms with two valence electrons, such as the alkaline-earth metals or ytterbium, possess ultra-narrow intercombination transitions from their singlet ground state to metastable triplet states. The ${^{1}\text{S}_{0}} - {^{3}\text{P}_{0}}$ transition, which connects two states that are free of electronic magnetic moment, is used as frequency reference in optical atomic clocks \cite{Ludlow2015oac} and is of interest for quantum simulation \cite{Martin2013qmb, Livi2016sda, Bromley2018doi, Cooper2015aco, Covey20182ri}, computation \cite{Daley2008qcw, Shibata2009asq, Gorshkov2009aem}, and gravitational wave detection \cite{Yu2011gwd, Graham2013nmf, Kolkowitz2016gwd}. The ${^{1}\text{S}_{0}} - {^{3}\text{P}_{2}}$ transition is equally narrow \cite{Porsev2004hfq}, but it connects the ground state to an excited state with electronic magnetic moment. This property has enabled high-resolution imaging of an Yb quantum gas in a magnetic field gradient \cite{Kato2012omr, Shibata2014osi}, a method that could also provide selective access to qubits in a quantum computer \cite{Daley2008qcw, Shibata2009asq}. Isotopes with nuclear spin exhibit hyperfine structure in the ${^{3}\text{P}_{2}}$ state, which will make it possible to induce nuclear spin state specific AC Stark shifts and Raman couplings using the ${^{1}\text{S}_{0}} - {^{3}\text{P}_{2}}$ transition. This property might allow the creation of artificial gauge fields that are significantly less hampered by off-resonant scattering of photons or collisions between metastable state atoms compared to schemes exploiting broader transitions \cite{Goldman2014lig, Lin2009bec, Mancini2015ooc, Songeaao2018oos,Leroux2018naa} or using metastable atoms \cite{Gerbier2010gff, Cooper2015aco, Livi2016sda}. Ultracold mixtures containing ${^{3}\text{P}_{2}}$ atoms have been obtained from quantum gases of ground state atoms by excitation on the ${^{1}\text{S}_{0}} - {^{3}\text{P}_{2}}$ transition, leading to the discovery of Feshbach resonances between Yb ${^{1}\text{S}_{0}}$ and ${^{3}\text{P}_{2}}$ atoms \cite{Kato2013cor}. These resonances are interesting for quantum information processing \cite{Daley2008qcw, Shibata2009asq, Gorshkov2009aem}, are predicted to show signatures of quantum chaos \cite{Green2016qci} and have been exploited to form Feshbach molecules \cite{Taie2016fre, Takasu2017moa}. Also mixtures of ${^{3}\text{P}_{2}}$ Yb with Li have been created \cite{Khramov2014uhm, Hara2014atd} and their collisional stability investigated \cite{Dowd2015mfd, Konishi2016cso, Gonzalez2013mtf, Petrov2015mco, Chen2015aif}. The ${^{1}\text{S}_{0}} - {^{3}\text{P}_{2}}$ transition might also be useful to induce optical Feshbach resonances \cite{Fedichev1996ion,Ciurylo2005oto, Takahashi2008ofr}, or to create quantum gases with quadrupole interactions \cite{Derevianko2001foc, Santra2004pom, Bhongale2013qpo, Lahrz2014dqi, Huang2014qpo}.

Many of these applications require quantum degenerate gases and so far three two-valence-electron elements have been cooled to quantum degeneracy: Yb \cite{Takasu2003ssb}, Ca \cite{Kraft2009bec}, and Sr \cite{Stellmer2009bec}. The frequency of the ${^{1}\text{S}_{0}} - {^{3}\text{P}_{2}}$ transition is only well known for Yb. Strontium has properties that significantly distinguish it from Yb, offering different opportunities. It enables higher phase space densities directly by laser cooling, which makes it possible to create quantum gases with large atom number or with high repetition rate \cite{Stellmer2013pqd}. Its fermionic isotope $^{87}$Sr has a nuclear spin of 9/2, which should enable better Pomeranchuck cooling \cite{Taie2012smi,Ozawa2018asc} or larger synthetic dimensions \cite{Boada2012qse}. In order to combine these favorable properties with the possibilities offered by the ${^{1}\text{S}_{0}} - {^{3}\text{P}_{2}}$ transition, the frequency of this transition needs to be determined to at least the MHz level.

In this article we report the measurement of the ultra-narrow $^{87}\text{Sr}$ ${^{1}\text{S}_{0}} - {^{3}\text{P}_{2}}$ transition by direct optical excitation. We perform loss spectroscopy of an ultracold strontium sample and determine the resonance frequency by comparison to four spectral lines of molecular iodine, which serves as a natural and documented reference. The iodine lines are identified by comparing a \si{\giga\hertz}-wide iodine spectrum around the Sr lines with the spectra calculated by the IodineSpec5 software~\cite{IodineSpec5}. The accuracy of the measurement is limited by the uncertainty in iodine transition frequencies, whereas the precision is limited by frequency drifts of an optical resonator used for spectroscopy laser stabilization. The relative frequency between the Sr transition and specific iodine lines is obtained with an accuracy of \SI{250}{\kilo\hertz} and the absolute frequency is limited by the iodine line accuracy of \SI{30}{\mega\hertz}. These measurements open the door to using the Sr ${^{1}\text{S}_{0}} - {^{3}\text{P}_{2}}$ transition for important applications, such as the creation of artificial gauge fields or quantum computation.

This article has the following structure: Sec.~\ref{sec:experimentalDetails} describes the spectroscopy laser system, the iodine spectroscopy setup, and Sr sample preparation; Sec.~\ref{sec:SrSpectroscopy} introduces relevant Sr transitions, presents initial coarse and final precise determination of the ${^{1}\text{S}_{0}} - {^{3}\text{P}_{2}}$ transition frequency and analyses the measurement error. Conclusions are given in Sec.~\ref{sec:conclusion}.

\section{Experimental details} \label{sec:experimentalDetails}

\subsection{${^1\text{S}_0} - {^3\text{P}_2}$  spectroscopy laser setup}

\begin{figure}[tb]
	\centering
	\includegraphics[width=0.99 \columnwidth]{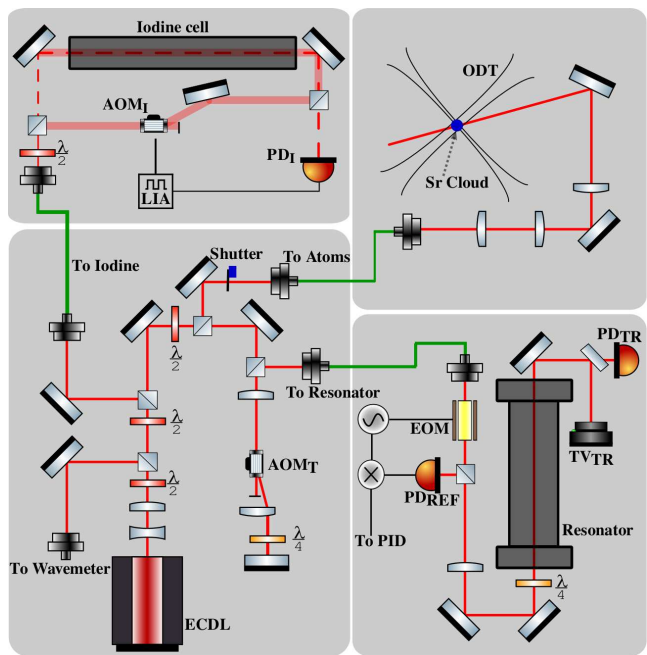}
	\caption{Scheme of the spectroscopy setup. The light of an extended cavity diode laser (ECDL) is distributed to a wavemeter, an optical resonator, an iodine spectroscopy setup, and Sr samples in an optical dipole trap (ODT). The laser frequency is stabilized to a mode of an optical resonator. Acousto-optic modulator $\text{AOM}_\text{T}$ introduces a controlled offset between the laser light and the resonator mode and is used to perform spectroscopy scans. To lock the light to the resonator the Pound-Drever-Hall method is used, for which an electro-optic modulator (EOM) creates sidebands on the light sent to the resonator and photodiode $\text{PD}_\text{REF}$ measures the light intensity reflected from it. The light transmitted through the resonator is analyzed by TV camera $\text{TV}_\text{TR}$ and photodiode $\text{PD}_\text{TR}$. The laser frequency can be referenced to iodine lines using the iodine spectroscopy setup, consisting of an iodine vapor cell, $\text{AOM}_\text{I}$, $\text{PD}_\text{I}$, and a lock-in amplifier (LIA).}
   \label{fig:lasersetup}
\end{figure}

Light for the spectroscopy of Sr and iodine is produced by an external cavity diode laser (ECDL; wavelength: 671\,nm; power: \SI{24}{\milli\watt}; diode: Toptica LD-0670-0035-AR-1), see Fig.~\ref{fig:lasersetup}. The ECDL is locked to a resonator (free spectral range \SI{1}{\giga\hertz}, length tunable by piezo) by the Pound-Drever-Hall (PDH) method~\cite{Black2001pdh}. A frequency shift of \SI{550}{\mega\hertz} to \SI{850}{\mega\hertz} is introduced by an AOM between the ECDL and the light used for locking. Spectroscopy scans are performed by slowly varying the AOM frequency so that the lock follows. The feedback loop uses a fast PID controller (Toptica FALC 110), providing feedback of \SI{1.9}{\mega\hertz} bandwidth to the ECDL current and \SI{10}{\kilo\hertz} bandwidth to the ECDL grating. Based on the error signal, we estimate the laser linewidth to be at most \SI{85}{\kilo\hertz}. The light is sent through polarization-maintaining single-mode optical fibers to the Sr sample and to the iodine spectroscopy setup. The absolute frequency of the spectroscopy laser can be obtained with a wavemeter (Toptica HighFinesse WSU-30, accuracy of 30\,MHz) calibrated to the ${^1\text{S}_0} - {^3\text{P}_1}$ transition frequency of $^{88}\text{Sr}$, known to within \SI{10}{\kilo\hertz}~\cite{Ferrari2003pfm}.

\subsection{Iodine spectroscopy setup} \label{IodineSetup}

Spectra of iodine molecule vapor (natural sample, essentially 100\% $^{127}\text{I}_2$~\cite{DeLaeter2003awe}) contained in a heated quartz cell are recorded using Doppler-free saturated absorption spectroscopy~\cite{Haensch1971hrs, Huang2013pfm}. We will now briefly describe the $\text{I}_2$ spectroscopy setup, see Fig~\ref{fig:lasersetup}. The quartz cell is \SI{60}{\centi\m} long and kept at approximately \SI{530}{\degreeCelsius} (not stabilized by feedback) in order to populate the higher vibrational levels of the iodine molecule~\cite{Huang2013pfm}. A cold finger, stabilized at $\SI{20\pm 0.3}{\degreeCelsius}$, is used to set the iodine partial pressure. The collimated spectrocopy beam entering the setup (waist $\SI{0.7}{\milli\m}$, power $\SI{5}{\milli\watt}$) is split into a pump and a probe beam. The probe beam (power \SI{0.5}{\milli\watt}) is sent through the cell onto a photodiode that records the spectroscopy signal. The pump beam is frequency shifted by acousto-optic modulator AOM$_\text{I}$, after which it has approximately \SI{3}{\milli\watt} of power, and gets sent through the iodine cell in a counterpropagating manner with respect to the probe~\cite{EndnoteSpectroscopyAOM}. We enhance the weak Doppler-free signal by lock-in detection. We use AOM$_\text{I}$ to chop the pump beam at \SI{50}{\kilo\hertz} (square wave) and we demodulate the detected probe signal at that frequency on a lock-in amplifier (EG \& G Instruments Model 7265) using a time constant of \SI{100}{\milli\second}. A few things are worth mentioning about this approach: first of all, the lock-in method is necessary, because the bare Lamb dips in the Doppler spectrum are too weak to be seen directly with the available power; secondly, the method is quite forgiving in terms of the chopping frequency, and in particular, the chopping frequency can be increased if one wants to reduce the time constant of the lock-in for faster scans, at the expense of signal-to-noise~\cite{EndnoteChoppingFrequency}; thirdly, the method is forgiving to slight misalignment in the overlap of the pump and probe beams, imperfect collimation of the beams, and slight power fluctuations of pump and probe.

\subsection{Strontium sample preparation and spectroscopy principle} \label{SrPrepAndSpec}

Spectroscopy of the Sr ${^1\text{S}_0} - {^3\text{P}_2}$ transition is done on an ultracold cloud of $^{87}$Sr in an equal mixture of all nuclear spin states contained in an optical dipole trap (ODT).  The ODT consists of two horizontally propagating, linearly-polarized \SI{1064}{\nano\meter} beams crossing at right angles and having waists of approximately \SI{60}{\micro\meter} and \SI{95}{\micro\meter} and powers of \SI{2.2}{\watt} and \SI{1.2}{\watt}; the beams have a \SI{160}{\mega\hertz} frequency difference in order to avoid mutual interference. To prepare the sample, a magneto-optical trap (MOT) is loaded from a Zeeman-slowed atomic beam and then transferred into the ODT using the techniques described in Ref.~\cite{Stellmer2013pqd}. We obtain a cloud of $2\times 10^5$ Sr atoms at \SI{360}{\nano\K}, which has a $1/e$-width of $\sim$\SI{18}{\micro\m} in the vertical direction and $\sim$\SI{27}{\micro\m} in the horizontal direction. We reduce the residual magnetic field to less than \SI{30}{\milli\gauss} at the location of the atomic cloud. The Sr spectroscopy beam is focused to a waist of about \SI{60}{\micro\m} at the sample position. Spectroscopy is performed time sequentially and measures frequency dependent loss of ground state atoms. A sample is prepared, exposed to spectroscopy light, which leads to atom loss, and the remaining ground-state atom number is detected by absorption imaging on the ${^1\text{S}_0} - {^1\text{P}_1}$ transition.

\section{Determination of the ${^1\text{S}_0} - {^3\text{P}_2}$ transition frequency} \label{sec:SrSpectroscopy}

\begin{figure}[tb]
	\centering
	\includegraphics[width=.98\columnwidth]{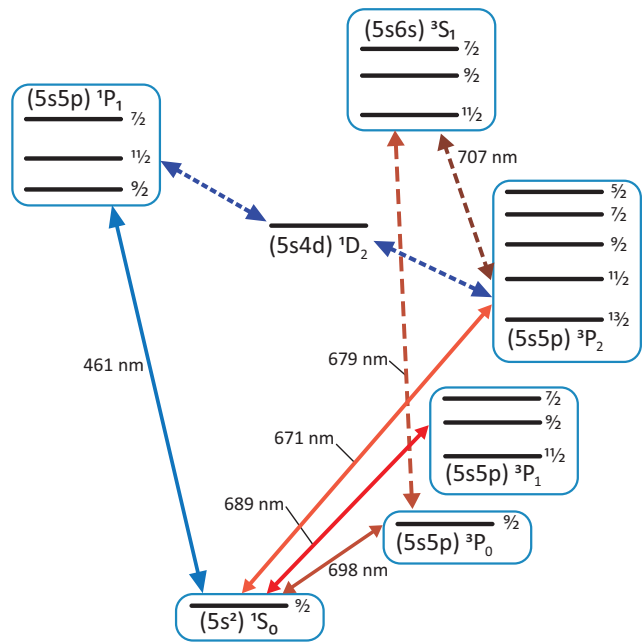}
 	\caption{\label{fig:SrLevelStructure} Level scheme of the low-lying electronic states of $^{87}\text{Sr}$. The transitions at \SI{461}{\nano\meter} and \SI{689}{\nano\meter} are used for MOTs, the transitions at \SI{679}{\nano\meter} and \SI{707}{\nano\meter} are repump transitions, the one at \SI{698}{\nano\meter} is the clock line, and the transition at \SI{671}{\nano\meter} is the ${^1\text{S}_0} - {^3\text{P}_2}$ line whose frequency we measure in this work. The hyperfine structure of the ${^1\text{D}_2}$ is neglected because it is irrelevant for this work.}
\end{figure}

Strontium levels and transitions that are relevant for this work are shown in Fig.~\ref{fig:SrLevelStructure}. The transitions ${^{1}\text{S}_{0}} - {^{3}\text{P}_{0, 2}}$ are dipole forbidden in isotopes with pure spin-orbit (LS) coupling because of spin and total angular momentum selection rules \cite{Boyd2007nse}. A small dipole matrix element can be induced by mixing  of the ${^{3}\text{P}_{0, 2}}$ states with ${^{1}\text{P}_{1}}$ through the application of a magnetic field or through hyperfine coupling in the case of $^{87}\text{Sr}$, the only stable Sr isotope with nuclear spin. For the bosonic ${^{88}\text{Sr}}$ the observation of the ${^1\text{S}_0} - {^3\text{P}_0}$ clock transition has been reported with an external mixing field as low as \SI{13}{\gauss}~\cite{Taichenachev2006mfi, Poli2009sol}. Most Sr optical lattice clocks use fermionic Sr in order to exploit hyperfine mixing to enable the clock transition.

In this work we use fermionic $^{87}\text{Sr}$, which allows dipole transitions between ${^1\text{S}_0}  ~(\text{F} = 9/2)$ and ${^3\text{P}_2}~(\text{F}^\prime = \{7/2,9/2,11/2\})$ with a linewidth of approximately \SI{1}{\milli\hertz} by hyperfine mixing \cite{Porsev2004hfq}. The transitions ${^1\text{S}_0}  ~(\text{F} = 9/2) - {^3\text{P}_2}~(\text{F}^\prime = \{5/2,13/2\})$ have $\Delta \text{F} = 2$ and the ground and excited states have opposite parity, which makes them dipole forbidden. They are however still accessible as magnetic quadrupole transitions (M2)~\cite{YamaguchiThesis} and we observe the ${^1\text{S}_0}  ~(\text{F} = 9/2) - {^3\text{P}_2}~(\text{F}^\prime = 5/2)$ transition.

We determine the ${^1\text{S}_0} - {^3\text{P}_2}$ transition frequency in two steps. The first step, described in Sec.~\ref{sec:CoarseIndirectMeasurement}, determines the transition indirectly to within $\sim$\SI{100}{\mega\hertz}. This is sufficiently precise to find the transition with direct spectroscopy, see Sec.~\ref{sec:precisionMeasurement}. In Sec.~\ref{sec:errorAnalysis} we discuss the error of our measurement.

\subsection{Coarse, indirect determination} \label{sec:CoarseIndirectMeasurement}

The ${^1\text{S}_0} - {^3\text{P}_2}$ transition frequency has only been measured for the most abundant isotope $^{88}$Sr with an accuracy of 120\,MHz \cite{Sansonetti2010wtp}. The $^{87}$Sr ${^1\text{S}_0} - {^3\text{P}_2}$ transition frequencies can be estimated by adding the $^{87}$Sr ${^3\text{P}_2}$ hyperfine shifts, which have been determined by radiofrequency spectroscopy in hot Sr~\cite{Heider1977hfs}, and the isotope shift. Here we assume that the ${^1\text{S}_0} - {^3\text{P}_2}$ isotope shift is the same as the measured ${^1\text{S}_0} - {^3\text{P}_0}$ and ${^1\text{S}_0} - {^3\text{P}_1}$ isotope shifts \cite{Bender1984isi, Buchinger1985iot, Takano2017pdo}, which are both within ~\SI{1}{\mega\hertz} of $f_{88}-f_{87} = $~\SI{62}{\mega\hertz}. We verify the estimated transition frequency by performing a simple, coarse and indirect frequency determination. We determine f(${^1\text{S}_0} - {^3\text{P}_2}$) using conservation of energy: we measure f(${^3\text{P}_2} - {^3\text{S}_1}$) and use the well-known transition frequencies f(${^1\text{S}_0} - {^3\text{P}_0}$) and f(${^3\text{P}_0} - {^3\text{S}_1}$) \cite{Courtillot2005aso} to calculate f(${^1\text{S}_0} - {^3\text{P}_2}$), see Fig.~\ref{fig:SrLevelStructure}. The ${^3\text{P}_2} - {^3\text{S}_1}$ transition is dipole-allowed, which makes it much broader and easier find than the doubly-forbidden mHz-linewidth ${^1\text{S}_0} - {^3\text{P}_2}$ transition. Similar schemes were used to determine the Sr ${^1\text{S}_0} - {^3\text{P}_0}$ transition \cite{Courtillot2005aso} and the Yb ${^1\text{S}_0} - {^3\text{P}_2}$ transition \cite{Yamaguchi2010hrl}.

To determine $f({^3\text{P}_2} - {^3\text{S}_1})$ we use reservoir spectroscopy \cite{Stellmer2014rst}. This technique relies on the fact that atoms in the ${^1\text{S}_0} - {^1\text{P}_1}$ MOT cycle can decay through the ${^1\text{D}_2}$ state into the metastable and magnetic ${^3\text{P}_2}$ state, the low field seeking $m_\text{F}$ substates of which are captured in the magnetic quadrupole field of the MOT. These atoms can be pumped back into the MOT cycle with light on a transition from the ${^3\text{P}_2}$ state to some higher-lying state that has a high chance of decaying into the ground state. We use the ${^3\text{S}_1}$ state as the higher-lying state, from which atoms decay to the ground state through the short-lived ${^3\text{P}_1}$ state. The ${^3\text{P}_2} - {^3\text{S}_1}$ transition corresponds to a repump laser operating around \SI{707}{\nano\meter}. Thus, when the repump laser is tuned to a resonance, the ${^3\text{P}_2}$ atoms from the magnetically trapped reservoir are quickly brought back into the ${^1\text{S}_0} - {^1\text{P}_1}$ MOT cycle, rapidly increasing the number of ground state atoms and causing a MOT fluorescence flash. There are nine repump resonances due to the number of hyperfine states in both ${^3\text{P}_2}$ and ${^3\text{S}_1}$, but we do not need to measure all of those transitions to determine $f({^3\text{P}_2} - {^3\text{S}_1})$.

We observe three ${^3\text{P}_2} - {^3\text{S}_1}$ repump resonances, which we can attribute to specific transitions between hyperfine states in the ${^3\text{P}_2}$ and ${^3\text{S}_1}$ manifolds using knowledge of the ${^3\text{P}_2}$ hyperfine structure~\cite{Heider1977hfs}, knowledge of the ${^3\text{S}_1}$ hyperfine structure~\cite{Courtillot2005aso}, and selection rules. The absolute frequencies of these transitions are obtained with the wavemeter that is also part of the ${^{1}\text{S}_{0}} - {^{3}\text{P}_{2}}$ spectroscopy setup. These measurements, combined with the known transition frequencies $f({^1\text{S}_0} - {^3\text{P}_0})$ and $f({^3\text{P}_0} - {^3\text{S}_1})$~\cite{Courtillot2005aso}, provides estimates of the individual transition frequencies $f({^1\text{S}_0}  ~(\text{F} = 9/2) - {^3\text{P}_2}~(\text{F}^\prime = \{7/2,9/2,11/2\}))$. This determination has an accuracy of $\sim$\SI{100}{\mega\hertz} and confirms the estimated transition frequencies. Using this good starting point we now expect to find the transitions quickly in a direct spectroscopy search.

\begin{figure}[tb]
	\centering
    \offinterlineskip
	\includegraphics[width=1\columnwidth]{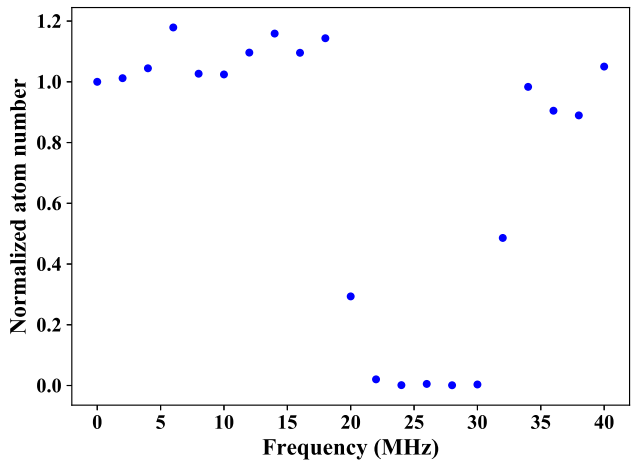}
	% Figure is made with Q:\groups\strontium\SrMicroscope\papers\1S0-3P2 spectroscopy\PaperFigures\DataAnalysis5_paper_Broadline.py
	\caption{\label{fig:exampleBroadTransition} Spectrum of the ${^1\text{S}_0}  ~(\text{F} = 9/2) - {^3\text{P}_2}~(\text{F}' = 11/2)$ transition broadened by using a large Rabi frequency (spectroscopy laser power of \SI{4.5}{\milli\watt}). The zero of the frequency axis is chosen arbitrarily.}
\end{figure}

\subsection{Precise, direct determination} \label{sec:precisionMeasurement}

We perform direct spectroscopy of the ${^1\text{S}_0} - {^3\text{P}_2}$ transitions using Sr samples in an ODT. When the spectroscopy laser is tuned into resonance, ground-state atoms are excited to the ${^3\text{P}_2}$ state, making them transparent to the ${^1\text{S}_0} - {^1\text{P}_1}$ absorption imaging beam. Return of excited state atoms to the ground state while remaining trapped is improbable since atoms in the ${^3\text{P}_2}$ state are not trapped in the 1064-\si{\nano\m} ODT and inelastic collisions involving ${^3\text{P}_2}$ atoms also lead to loss from the trap. To find a resonance we measure the fraction of ${^1\text{S}_0}$ atoms remaining in the ODT as a function of spectroscopy laser frequency, while keeping other parameters, such as illumination time and laser power, constant. During the first search for the transition we use the full power of the spectroscopy beam (\SI{4.5}{\milli\watt}). We repeatedly prepare Sr samples and use each to scan a \SI{1}{\mega\hertz} frequency interval over \SI{1}{\second}. An example for the resulting spectrum is shown in Fig.~\ref{fig:exampleBroadTransition} and determines all ${^1\text{S}_0} - {^3\text{P}_2}$ transitions to within \SI{10}{\mega\hertz}, using the known ${^3\text{P}_2}$ hyperfine splittings.

\begin{figure}[tb]
	\centering
	\includegraphics[width=.98\columnwidth]{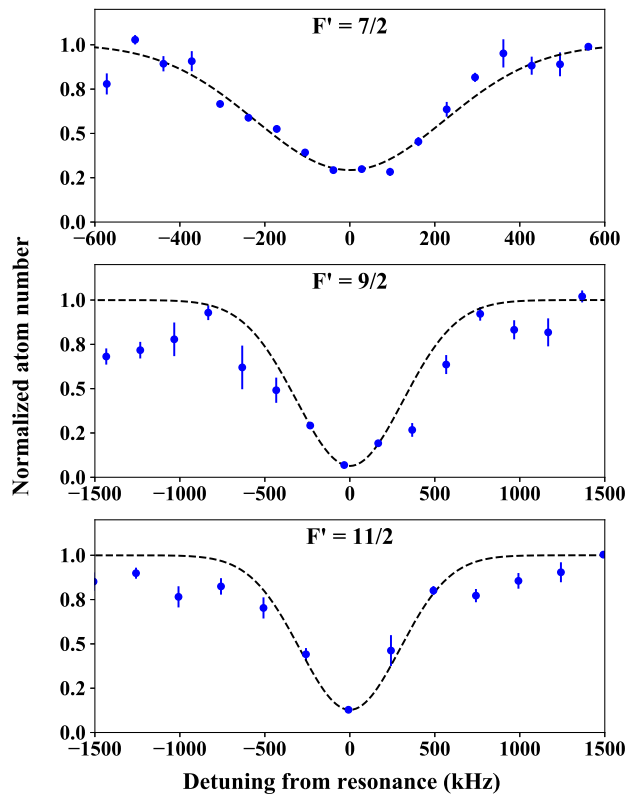}
	% Figure is made with Q:\groups\strontium\SrMicroscope\papers\1S0-3P2 spectroscopy\PaperFigures\DataAnalysis5_paper.py
	\caption{\label{fig:narrow3P2lines} High-resolution spectra of the $^1\text{S}_0 - {^3\text{P}_2}$ ($\text{F}' = \{7/2,9/2,11/2\}$) transition in $^{87}\text{Sr}$ measured by atom loss spectroscopy in an ODT using low spectroscopy beam power (\SI{0.45}{\milli\watt}). The atom numbers are normalized to the ones far away from any spectroscopy signal. The error bars represent the standard error of five measurements per data point.}
\end{figure}

\begin{table}[b]
	\caption{Frequencies and measured linewidths of the ${^1\text{S}_0} - {^3\text{P}_2}$ transition to three different hyperfine states in the ${^3\text{P}_2}$ manifold. The frequencies are determined using a wavemeter, whereas the full-width half-maximum (FWHM) linewidths are obtained from Gaussian fits to the spectroscopy signals shown in Fig.~\ref{fig:narrow3P2lines}.}
    % One obtains these results using the script
    % Q:\groups\strontium\SrMicroscope\papers\1S0-3P2 spectroscopy\PaperFigures\DataAnalysis5_paper.py
    \begin{ruledtabular}
		\begin{tabular}{c c c  }
			$\text{F}'$ & frequency & FWHM linewidth\\
            & (\si{\mega\hertz}) & (\si{\mega\hertz})\\
            \hline \\
            $7/2$  & $\SI{446648769\pm 30}{}$ & 0.52(4)\\
            $9/2$  & $\SI{446647793\pm 30}{}$ & 0.74(3)\\
            $11/2$  & $\SI{446646618\pm 30}{}$ & 0.69(8)\\
		\end{tabular}
	\end{ruledtabular}
	\label{tab:linefreqs}
\end{table}

\begin{figure*}[tb]
	\centering
	\includegraphics[width=1.0\textwidth]{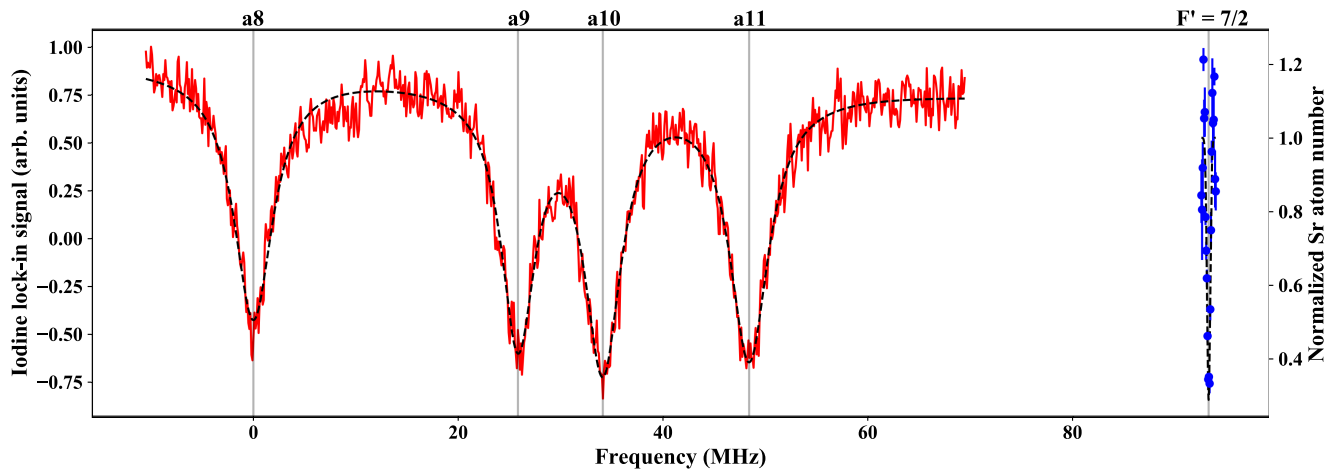}
	%Figure made with Q:\groups\strontium\SrMicroscope\papers\1S0-3P2 spectroscopy\PaperFigures\DatasetAnalysis_paper_combinedSpec.py
	\caption{\label{fig:IodineAnd7halftogether}  (Color online) Combined spectrum of the $^{87}\text{Sr}$ ${^1\text{S}_0} - {^3\text{P}_2}$ $(\text{F}' = 7/2)$ line and iodine lines \{a8, a9, a10, a11\}. The red line is iodine data, the blue circles are $^{87}\text{Sr}$ data, the black dashed lines are the fits to the respective spectra (multiple Lorentzian for iodine and Gaussian for $^{87}\text{Sr}$). The center of each fitted spectral line is marked with a vertical gray line and labeled on top. The center of the a8 line is chosen as the zero of the frequency axis.}
\end{figure*}

In order to determine the transition frequency with more precision we zoom in on the detected broad spectroscopy signals by recording spectra with \SI{0.45}{\milli\watt} of spectroscopy laser power and reduced frequency interval and illumination time per sample. The narrowest spectroscopy signals for $\text{F}' = \{7/2,9/2,11/2\}$ are shown in Fig.~\ref{fig:narrow3P2lines} together with Gaussian fits, the parameters of which are summarized in Table~\ref{tab:linefreqs}. The values of the transition frequencies measured with the wavemeter are also reported in Table~\ref{tab:linefreqs}, and the uncertainty is dominated by wavemeter inaccuracy.

Next we determine the frequency difference between four iodine transitions and the $^{87}\text{Sr}$ ${^1\text{S}_0}  ~(\text{F} = 9/2) - {^3\text{P}_2}~(\text{F}' = 7/2)$ transition. This allows us to determine the frequency of the Sr transition with the accuracy of the known iodine transition frequencies, which currently is the same as the accuracy of the wavemeter (\SI{30}{\mega\hertz}), but can be improved in the future with iodine spectroscopy alone \cite{Huang2013pfm}. More importantly the accuracy of the frequency difference (\SI{250}{\kilo\hertz}) is much higher than the accuracy of the absolute frequency, which  makes it possible to find the Sr transition with simple iodine spectroscopy and to lock the Sr laser to iodine lines.

The $\text{F}' = 7/2$ state is used for iodine comparison because it is within the spectroscopy AOM tuning range of the strong iodine transitions  $(J'-J'' = 32-33)(\nu'-\nu'' = 9-9)$ \{a8, a9, a10, a11\}~\cite{EndnoteIodineNotation}. The frequencies corresponding to the other hyperfine states in the ${^3\text{P}_2}$ manifold can be found by using this frequency and the known ${^3\text{P}_2}$ hyperfine splittings~\cite{Heider1977hfs}. Figure~\ref{fig:IodineAnd7halftogether} presents an example of a recorded iodine spectrum, fitted with Lorentzians, combined with the Sr spectrum, fitted with a Gaussian. Table~\ref{tab:iodinefreqsand7half} lists the fitted central frequency values of the iodine transitions, where each value is an averaged result from the fits to four measured spectra. Using the a8 frequency calculated by IodineSpec5 we obtain \SI{446648775 \pm 30}{\mega\hertz} for the $^{87}\text{Sr}$ ${^1\text{S}_0}  ~(\text{F} = 9/2) - {^3\text{P}_2}~(\text{F}' = 7/2)$ transition frequency, which is consistent with the frequency determined by the wavemeter.

\begin{table}[b]
	\caption{Measured frequencies of the iodine transitions \{a8, a9, a10, a11\} within the manifold P(33) (9-9) and the $^1\text{S}_0 - {^3\text{P}_2}$ ($\text{F}' = 7/2$) transition frequency. The frequencies are reported with respect to the measured frequency of the a8 transition. The iodine transition frequencies are given for our conditions of the iodine cell (see text) and are shifted by -100(15)\,kHz with respect to iodine lines at zero pressure and temperature. The error of the Sr transition is dominated by drifts of the reference resonator.}
	% Table made from data generated with Q:\groups\strontium\SrMicroscope\papers\1S0-3P2 spectroscopy\PaperFigures\Data180418\D1\DatasetAnalysis_paper.py python script
	\begin{ruledtabular}
		\begin{tabular}{l  c  }
			transition & frequency\\
            & (\si{\mega\hertz})\\
            \hline \\
			a8  &  \SI{0.00 \pm 0.03}{} \\
			a9  &  \SI{25.88 \pm 0.03}{}  \\
			a10  &  \SI{34.17 \pm 0.05}{} \\
			a11  &  \SI{48.39 \pm 0.04}{} \\
            $^1\text{S}_0 - {^3\text{P}_2}$ ($\text{F}' = 7/2$)  & \SI{93.27 \pm 0.25}{} \\
		\end{tabular}
	\end{ruledtabular}
	\label{tab:iodinefreqsand7half}
\end{table}

\begin{figure}[tb]
	\centering
	\includegraphics[width=.98\columnwidth]{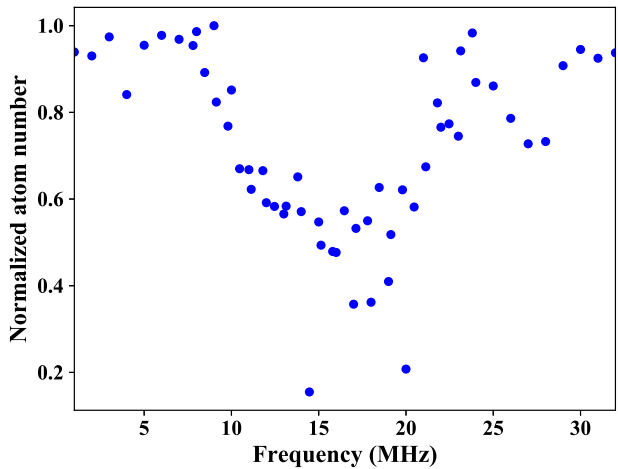}
	% Figure is made with Q:\groups\strontium\SrMicroscope\papers\1S0-3P2 spectroscopy\PaperFigures\DataAnalysis6.py
	\caption{\label{fig:5halfline} Spectrum of the ${^1\text{S}_0}  ~(\text{F} = 9/2) - {^3\text{P}_2}~(\text{F}' = 5/2)$ M2 transition with a spectroscopy beam power of \SI{4.5}{\milli\watt}. The zero of the frequency axis is chosen arbitrarily.}
\end{figure}

The frequency of the $^1\text{S}_0 - {^3\text{P}_2}$ ($\text{F}' = 5/2, 13/2$) M2 transitions can be determined from the previous measurement and the ${^3\text{P}_2}$ hyperfine splittings. Guided by this calculation we observe the $^1\text{S}_0 - {^3\text{P}_2}$ ($\text{F}' = 5/2$) transition by direct optical excitation. Since it is an M2 transition it is expected to be much weaker than the HFM-E1 lines, and we indeed must use about 10 times larger intensity and a 20 times longer illumination time to induce observable atom loss on this transition compared to the case of the dipole transitions. Fig.~\ref{fig:5halfline} shows a spectrum of this line recorded with a spectroscopy beam power of \SI{4.5}{\milli\watt} and an illumination time of \SI{10}{\second}.

Based on our measurement of the ${^1\text{S}_0} - {^3\text{P}_2}$ transition and the previously reported results for the ${^1\text{S}_0} - {^3\text{P}_1}$ and ${^3\text{P}_1} - {^3\text{S}_1}$ transition frequencies and hyperfine splittings of all mentioned states \cite{Courtillot2005aso, Heider1977hfs}, we can also give a more accurate value for the ${^3\text{P}_2} (F = 7/2) - {^3\text{S}_1} (F' = 7/2)$ repumping transition frequency, which evaluates to \SI{423914969 \pm 30}{\mega\hertz}. The corresponding values involving any other hyperfine states can be easily calculated from the known hyperfine splittings \cite{Courtillot2005aso,Heider1977hfs}.

\subsection{Error analysis} \label{sec:errorAnalysis}

We measure absolute frequencies in two ways, using the wavemeter or using iodine lines as reference \cite{IodineSpec5}. Both methods have an uncertainty of 30\,MHz, which dominates all other sources of error. We also determine the relative frequency between the Sr transitions and the iodine lines. The error in the relative frequency is much smaller than the absolute error and will be discussed in the following.

The error in the relative frequency measurement has statistical and systematic components. Statistical errors arise from the drifts of the Fabry-Perot resonator to which the laser is locked, from errors in the fits used to determine the center of spectral lines, and from iodine line shifts by iodine temperature and pressure changes~\cite{Dareau2015dsy, EndnoteAOMdrift}. Systematic errors are the collisional shifts of the iodine lines, which effectively move our frequency reference point from its literature value~\cite{Dareau2015dsy}, and the AC Stark shift of the Sr transitions by the dipole trap light.

The dominant contribution to the statistical error comes from changes of the cavity resonance to which the spectroscopy laser is locked while the measurements are performed. This error is estimated by recording iodine spectra several times before, during, and after the one-hour timespan during which the Sr spectra are recorded, and then analyzing the drift of the relative frequency between the cavity resonance and the iodine lines. We assume that the iodine lines do not change significantly over this timespan (the validity of that assumption will be analyzed below), therefore the drift is due to changes of the resonator frequency. The maximum drift we observe is~\SI{250}{\kilo\hertz}.

Errors also originate from pressure and temperature shifts of the iodine lines. Compared to a zero temperature and pressure gas the iodine lines are shifted by $\delta f_{I_2} = \alpha_S P_{\text{I}_2} T^{-7/10}$ at pressure $P$ and temperature $T$, where $\alpha_S=\SI{-400\pm 60}{\kilo\hertz~ \K^{7/10}\per\pascal}$ is an empirically determined proportionality constant~\cite{Dareau2015dsy}. $P_{\text{I}_2}$ is set by the cold finger temperature, and the relevant iodine vapor pressure equation is given in Ref.~\cite{Huang2013pfm}, whereas $T$ is set by the iodine cell body temperature. Statistical errors arise from uncertainties in $P$ and $T$. The cold finger temperature uncertainty of \SI{0.3}{\K} translates into a pressure uncertainty below \SI{1}{\pascal}. The body temperature has an uncertainty below \SI{20}{\K}. These uncertainties lead to a statistical error of \SI{3}{\kilo\hertz} in the iodine line frequency. The systematic shift of the iodine lines is $\delta f_{I_2} = \SI{-100(15)}{\kilo\hertz}$, where the dominant contribution to the error arises from uncertainties in $\alpha_S$. The values reported in Table \ref{tab:iodinefreqsand7half} are given in presence of this shift.

Another source of error is the light shift induced on the Sr transition by the ODT. We obtain an upper limit for this shift by recording spectra using ODT depths up to a factor two higher or lower than the depth used usually. We do not observe a correlation of the transition frequency with the ODT depth, which we attribute to drifts of the reference resonator during the few hours that we spent to record this data. We conclude that the light shifts are insignificant compared to the resonator drifts of~\SI{250}{\kilo\hertz}. The Sr spectroscopy lines are broadened by ODT light shifts, the Doppler effect, Zeeman shifts of the unresolved $m_F$ levels, and collisional effects.

\section{Conclusion} \label{sec:conclusion}

We have determined the frequency of the $^{87}\text{Sr}$ $^1\text{S}_0 - {^3\text{P}_2}$ transition with an accuracy of
\SI{30}{\mega\hertz} and the frequency difference of that transition to molecular iodine lines with an accuracy of \SI{250}{\kilo\hertz}. This knowledge enables the use of simple iodine spectroscopy to find the $^{87}\text{Sr}$ $^1\text{S}_0 - {^3\text{P}_2}$ transition frequency or to lock a Sr $^1\text{S}_0 - {^3\text{P}_2}$ laser. Our work prepares the use of this Sr transition for applications, such as quantum simulation or computation.

\section{Acknowledgements} \label{sec:ack}

We thank Jan Matthijssen for technical assistance in the early stages of the project and H. Kn\"{o}ckel, B. Bodermann, and E. Tiemann for the software IodineSpec5 \cite{IodineSpec5}. This project has received funding from the European Research Council (ERC) under the European Union's Seventh Framework Programme (FP7/2007-2013) (Grant agreement No. 615117 QuantStro). We thank the Netherlands Organisation for Scientific Research (NWO) for funding through Vici grant No. 680-47-619 and Gravitation grant No. 024.003.037, Quantum Software Consortium. G. S. thanks the European Commission for Marie Curie grant SYMULGAS, No. 661171. C.-C. C. thanks the Ministry of Education of the Republic of China (Taiwan) for a MOE Technologies Incubation Scholarship.

%\bibliography{1S03P2spectroscopy}

\begin{thebibliography}{71}
\bibitem [{End({\natexlab{c}})}]{EndnoteIodineNotation}%
	\BibitemOpen
	\href@noop {} {}\bibinfo {note} {{Here, $J''$ and
			$\nu''$ denote the lower rotational and vibrational levels respectively, and
			$J', \nu'$ denote the upper level. The symbols \{a8, a9, a10, a11\} label the
			hyperfine states~\cite{IodineSpec5}. The common designation for these
			rovibrational transitions is P(33) (9-9).}}\BibitemShut {Stop}%
\bibitem [{End({\natexlab{d}})}]{EndnoteAOMdrift}%
	\BibitemOpen
	\href@noop {} {} \bibinfo {note} {{The AOM frequency
			uncertainty is negligible compared to other error sources as it is driven by
			a direct digital synthesizer synchronized to a rubidium atomic
			clock.}}\BibitemShut {Stop}%
\end{thebibliography}


\end{document}